\documentclass[%
aip,
 amsmath,amssymb,
 reprint,%
tightenlines,
flushbottom,
]{revtex4-1}

\usepackage{graphicx}
\usepackage{dcolumn}
\usepackage{bm}

\usepackage{booktabs}
\usepackage{ wasysym }
\usepackage{siunitx}
\usepackage[utf8]{inputenc}
\usepackage[T1]{fontenc}
\usepackage{mathptmx}
\usepackage{etoolbox}
\usepackage{hyperref}
\hypersetup{
    colorlinks = true,
    linkcolor = blue,
    citecolor = blue,
    urlcolor=blue,
}

\usepackage{tabularx}

\makeatletter
\def\@email#1#2{%
 \endgroup
 \patchcmd{\titleblock@produce}
  {\frontmatter@RRAPformat}
  {\frontmatter@RRAPformat{\produce@RRAP{*#1\href{mailto:#2}{#2}}}\frontmatter@RRAPformat}
  {}{}
}%
\makeatother

\usepackage{setspace}
  \usepackage{parskip}
\usepackage[compact]{titlesec}         
\titlespacing{\section}{4pt}{4pt}{4pt} 
\AtBeginDocument{
  \setlength\abovedisplayskip{4pt}
  \setlength\belowdisplayskip{4pt}
  }
  
  \usepackage[]{caption, subfig}

\begin{document}

\preprint{AIP/123-QED}

\title{Plasma flows during the ablation stage of an over-massed pulsed-power-driven exploding planar wire array}
\author{R. Datta}
\affiliation{ Plasma Science \& Fusion Center, Massachusetts Institute of Technology, MA 02139, Cambridge, USA\looseness=-1 
}%
\author{J. Angel}
\affiliation{ 
Laboratory of Plasma Studies, Cornell University, Ithaca, NY, USA 
}%
\author{J.B. Greenly}
\affiliation{ 
Laboratory of Plasma Studies, Cornell University, Ithaca, NY, USA 
}%
\author{S.N. Bland}
\affiliation{ 
Blackett Laboratory, Imperial College London, London SW7 2BW, United Kingdom 
}%
\author{J.P. Chittenden}
\affiliation{ 
Blackett Laboratory, Imperial College London, London SW7 2BW, United Kingdom 
}%
\author{E. S. Lavine}
\affiliation{ 
Laboratory of Plasma Studies, Cornell University, Ithaca, NY, USA 
}%
\author{W. M. Potter}
\affiliation{ 
Laboratory of Plasma Studies, Cornell University, Ithaca, NY, USA 
}%
\author{D. Robinson}
\affiliation{ Plasma Science \& Fusion Center, Massachusetts Institute of Technology, MA 02139, Cambridge, USA\looseness=-1 
}%
\author{T. W. O. Varnish}
\affiliation{ Plasma Science \& Fusion Center, Massachusetts Institute of Technology, MA 02139, Cambridge, USA\looseness=-1 
}%
\author{E. Wong}
\affiliation{ Plasma Science \& Fusion Center, Massachusetts Institute of Technology, MA 02139, Cambridge, USA\looseness=-1 
}%

\author{D. A. Hammer}
\affiliation{ 
Laboratory of Plasma Studies, Cornell University, Ithaca, NY, USA 
}%
\author{B.R. Kusse}
\affiliation{ 
Laboratory of Plasma Studies, Cornell University, Ithaca, NY, USA 
}%
\author{J.D. Hare*}%
\email{jdhare@mit.edu.}
\affiliation{ Plasma Science \& Fusion Center, Massachusetts Institute of Technology, MA 02139, Cambridge, USA\looseness=-1 
}%


\date{\today}

\begin{abstract}
 We characterize the plasma flows generated during the ablation stage of an over-massed exploding planar wire array, fielded on the COBRA pulsed-power facility (1 MA peak current, 250 ns rise time). The planar wire array is designed to provide a driving magnetic field ($80-100$ T) and current per wire distribution (about $60$ kA), similar to that in a 10 MA cylindrical exploding wire array fielded on the Z machine. Over-massing the arrays enables continuous plasma ablation over the duration of the experiment. The requirement to over-mass on the Z machine necessitates wires with diameters of $75- \qty{100}{\micro \meter}$, which are thicker than wires usually fielded on wire array experiments. To test ablation with thicker wires, we perform a parametric study by varying the initial wire diameter between $33-\qty{100}{\micro \meter}$. The largest wire diameter (\qty{100}{\micro \meter}) array exhibits early closure of the AK gap, while the gap remains open during the duration of the experiment for wire diameters between $33-\qty{75}{\micro \meter}$. Laser plasma interferometry and time-gated XUV imaging are used to probe the plasma flows ablating from the wires. The plasma flows from the wires converge to generate a pinch, which appears as a fast-moving  ($V \approx \qty{100}{\kilo \meter \per \second}$) column of increased plasma density ($\bar{n}_e \approx \qty{2e18}{\per \centi \meter \cubed}$) and strong XUV emission. Finally, we compare the results with three-dimensional resistive-magnetohydrodynamic (MHD) simulations performed using the code GORGON, the results of which reproduce the dynamics of the experiment reasonably well.

\end{abstract}

\maketitle

\newcommand{\ra}[1]{\renewcommand{\arraystretch}{#1}}

\setlength{\textfloatsep}{4pt plus 0.5pt minus 0.5pt}

\section{\label{sec:intro} Introduction}

Inverse (or ``exploding'') cylindrical wire arrays are a commonly-used pulsed-power-driven source of magnetized plasma for laboratory astrophysics applications. These arrays consist of a cylindrical cage of thin conducting wires surrounding a central cathode. This magnetic field configuration drives radially-diverging outflows into a vacuum region, providing good diagnostic access.\cite{Harvey-Thompson2009} These arrays have previously been fielded on 1-MA university-scale facilities to study a variety of astrophysical phenomena, including magnetized plasma shocks,\citep{lebedev2014formation,burdiak2017structure,russell2022perpendicular,datta2022structure} laboratory magnetospheres,\citep{suttle2019interactions} and magnetic reconnection.\cite{hare2018experimental,suttle2018ion} For such applications, the wire arrays are typically over-massed, so that they provide continuous sustained plasma flows over the duration of the experiment.\citep{lebedev2002snowplow}

On larger pulsed-power machines, such as the Z machine (30 MA peak current, Sandia National Labs), \cite{sinars2020review,gomez2014magnetic} over-massed exploding wire arrays require a larger initial mass due to the higher driving current.\citep{lebedev2002snowplow} For the same wire material, this necessitates more wires and/or the use of larger diameter wires. Although arrays with thin  ($5-\qty{40}{\micro \meter}$ diameter) wires have been characterized extensively in pulsed-power-driven experiments,\citep{lebedev2002snowplow,shelkovenko2007wire,lebedev2019exploring} there has been little systematic effort to study ablation from thick ($>\qty{50}{\micro \meter}$ diameter) wires, especially with Z-relevant $>100$ T driving magnetic fields.\citep{gomez2014magnetic}

In cylindrical wire arrays, the maximum driving magnetic pressure is limited by the size of the central cathode and the AK gap (the gap between the anode/wires and cathode). The driving magnetic field in a cylindrical array can be determined from Ampere's law: $B(t) = \mu_0 I(t) / (2 \pi R)$, which shows that it varies inversely with the radius $R$ of the array. \citep{lebedev2002snowplow} 
For a $R = \qty{10}{\milli \meter}$ array, the peak driving field on a typical 1-MA university-scale machine is $B = 20$ T. The finite size of the central cathode makes it difficult to achieve $\sim 100$ T driving magnetic fields using cylindrical arrays on 1-MA university-scale facilities. In order to overcome this limitation, we explore the use of planar wire arrays to test ablation from thick wires in Z-relevant driving magnetic fields.  

These planar wire arrays consist of a linear arrangement of wires separated by a small AK gap from a planar return electrode. This ``exploding'' planar geometry, which has previously been fielded on 1-MA pulsed-power devices,\citep{bland2004use} allows us to achieve higher driving magnetic fields than in cylindrical arrays. 
We also investigate the use of planar wire arrays as a platform for laboratory astrophysics experiments.
Since exploding cylindrical arrays generate azimuthally symmetric flows, the majority of the plasma (and therefore the stored energy) is lost in directions that are not of interest.\citep{hare2018experimental,suttle2018ion} Moreover, due to radially-diverging flows, the density and advected magnetic field decrease rapidly with distance from the wires.\citep{burdiak2017structure,datta2022structure} 
In contrast, planar wire arrays could provide directed flows of denser plasma with higher advected magnetic fields, which can be desirable for many laboratory astrophysics applications. In magnetic reconnection experiments, for instance, this would increase dissipation in the current sheet, which is necessary for studying radiative-cooling effects.\cite{hare2022simulations,datta2022plasma}

 \begin{figure*}
\includegraphics[page=1,width=1.0\textwidth]{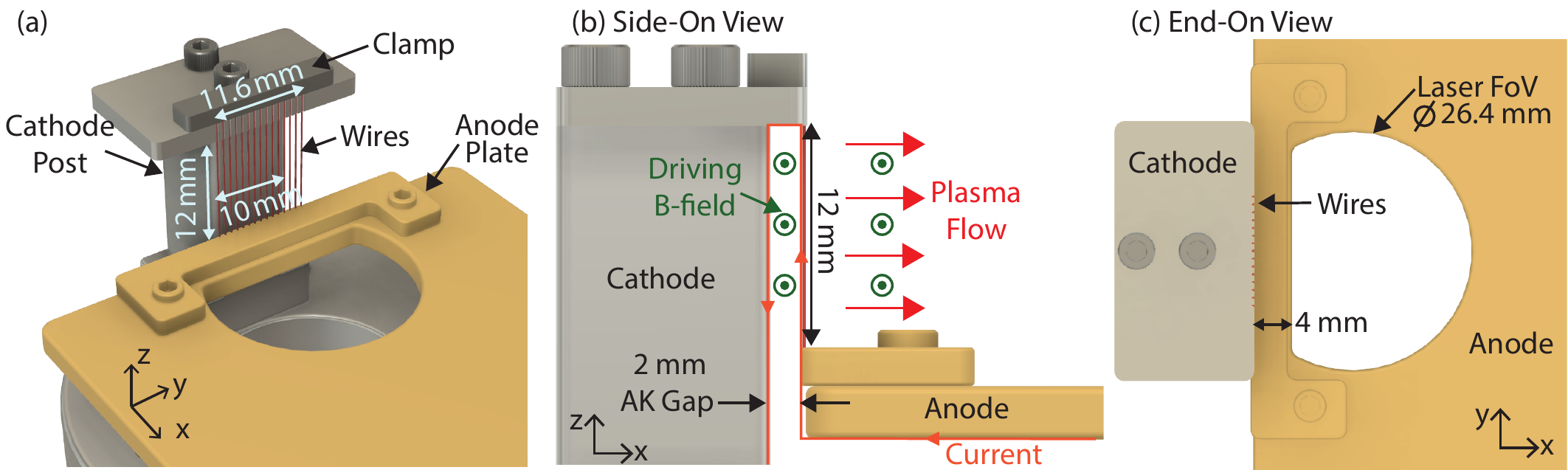}
\centering
\caption{(a) 3D CAD representation of the load hardware. The load hardware consists of a planar array of 15 equally-spaced aluminum wires.
(b) Side-on view of the load hardware.
(c) End-on view of the load hardware.
}
\label{fig:load}
\end{figure*}

Planar wire arrays were primarily developed as an efficient X-ray radiation source for indirect-drive inertial confinement fusion (ICF) experiments. \cite{kantsyrev2006planar,esaulov2006wire,kantsyrev2007properties,kantsyrev2008double,jones2010planar} In contrast to the ``exploding'' geometry used in this paper, the wire arrays used for X-ray generation typically consist of a linear row of wires between the cathode and anode of the pulsed-power device, without the planar return electrode placed adjacent to the wires. Furthermore, these arrays use thin $5-\qty{20}{\micro \meter}$ diameter wires, which implode during the course of the experiment.\cite{kantsyrev2006planar,kantsyrev2007properties,kantsyrev2008double} The arrays are designed such that the implosion time matches the time of peak current, in order to maximize X-ray emission (hence, they are also called matched arrays).\citep{kantsyrev2008double} The implosion stage typically proceeds in a cascade-like fashion, where the imploding wires, starting from the outermost wires, accelerate toward the geometric center of the array, to form a strongly-radiating inhomogeneous plasma column.\citep{esaulov2006wire,kantsyrev2008double} These array configurations have been reported to exhibit peak X-ray power and yield higher than imploding cylindrical arrays with similar number of wires.\citep{kantsyrev2007properties,kantsyrev2008double} \citeauthor{bland2004use} were the first to field the planar wire array in an exploding geometry. This geometry, which consisted of a matched planar array with thin $\qty{7.5}{\micro \meter}$ tungsten wires, exhibited a $5-6\times$ higher ablation rate compared to cylindrical wire arrays, consistent with the increased driving magnetic pressure inside the AK gap.\citep{bland2004use} Furthermore, the ablating plasma converged to form a magnetic precursor column offset from the plane of the wires, before exhibiting the cascade-like implosion described above.\citep{bland2004use} 

In contrast to the previous planar wire array experiments described above, which use matched arrays with thin wires, we use over-massed arrays with thick $33-100\qty{}{\micro \meter}$ aluminum wires. This suppresses the implosion stage, and generates continuous plasma ablation over the course of the experiment. In wire arrays, the initial flow of current through the wires forms dense cold wire cores surrounded by low-density coronal plasma. \cite{lebedev2002snowplow,shelkovenko2007wire}  Current density is concentrated in a thin skin region that surrounds the wire cores, and includes the coronal plasma.  The coronal plasma is redirected by the ${\bf j \times B}$ force of the driving magnetic field in the AK gap between the wires and the cathode. In a planar wire array, this generates plasma flows directed away from the AK gap.\cite{bland2004use} The ablating plasma advects some magnetic field from the AK gap as it flows outwards, creating outflows of magnetized plasma. In matched arrays, when the stationary wire cores begin to run out of mass (typically $\sim 50-80\%$ of the initial mass), periodic breaks appear in the wires, driven by the growth of a modified $m=0$-like axial instability. \cite{lebedev2002snowplow,chittenden2004equilibrium,jones2005study,knapp2010growth} This marks the end of the ablation phase, and the beginning of the implosion phase.\cite{lebedev2002snowplow} 

Large wire cores can be undesirable for the ablation process. A large wire core diameter relative to the inter-wire separation inhibits the ablation of mass and the advection of magnetic field with the ablating plasma. \cite{velikovich2002perfectly} A large core size may also increase the likelihood of AK gap closure in pulsed-power-driven systems. In wire arrays, closure of the AK gap is undesirable, as it short-circuits the current path, leading to decreased current flow through the wires and reduced/terminated ablation.  Previous experiments aimed at characterizing wire core size in imploding wire arrays show that core diameter varies with wire material and initial wire diameter, but is largely independent of the current per wire, and the inter-wire separation. \cite{shelkovenko2007wire}

In this paper, we explore the use of an over-massed exploding planar wire array as a platform for laboratory astrophysics experiments, and as a scaled experiment to investigate the ablation of thick wires in cylindrical wire arrays driven by Z-relevant driving magnetic fields. The array is driven by the COBRA pulsed-power machine (1 MA peak current, 250 ns rise time),\citep{greenly20081} and is designed to exhibit a magnetic driving pressure, current per wire, and inter-wire separation, comparable to that of a $\qty{40}{\milli \meter}$ diameter exploding wire array driven by a 10 MA current pulse from the Z machine.\cite{hare2022simulations,datta2022plasma} These experiments, therefore, allow us to investigate wire ablation on smaller 1 MA facilities, in loads designed for use on $\sim 10$ MA machines. We note that \citeauthor{bland2004use} also aimed to match the driving magnetic field of a 20 MA, 100 ns rise time current pulse on the Z Machine, to understand imploding cylindrical wire array ablation at higher current per wire and driving magnetic fields. In this paper, we target the driving conditions generated when Z operates in a synchronous long-pulse configuration, with 20 MA peak current (split between two arrays) and a 300 ns rise time.\citep{hare2022simulations,datta2022plasma} As such, we use the long-pulse mode on COBRA, as described in Sec. \ref{sec:setup}.

The requirement to over-mass on the Z machine necessitates wires with diameters of 75-\qty{100}{\micro \meter}, which are thicker than wires usually fielded on wire-array z-pinch experiments. To investigate ablation with thicker wires, we vary the initial wire diameter between $33-\qty{100}{\micro \meter}$ over multiple shots. The load hardware, as well as the magnetic field and current distributions in the load, are described in Sec. \ref{sec:load}. We characterize the plasma ablation and the reduction in the AK gap for the different wire sizes using laser shadowgraphy, Mach-Zehnder imaging interferometry, and XUV pinhole imaging (Sec. \ref{sec:diagnostic}). These experimental results are provided and discussed in Sec. \ref{sec:results} and Sec. \ref{sec:discussion}. Finally, in Sec. \ref{sec:sim}, we compare the experimental results with three-dimensional resistive magnetohydrodynamic (MHD) simulations performed using GORGON.

 \begin{figure}
\includegraphics[page=11,width=0.48\textwidth]{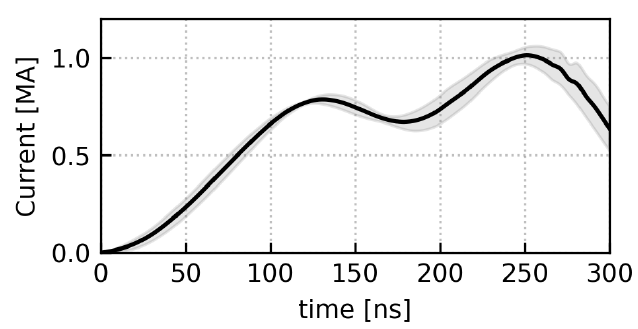}
\centering
\caption{Variation of the current delivered with time for the COBRA generator. We show the current pulse averaged over 6 successive shots. The shaded region is the shot-to-shot deviation in the current delivered.
}
\label{fig:current_pulse}
\end{figure}

 \begin{figure*}
\includegraphics[width=1.0\textwidth,page=2]{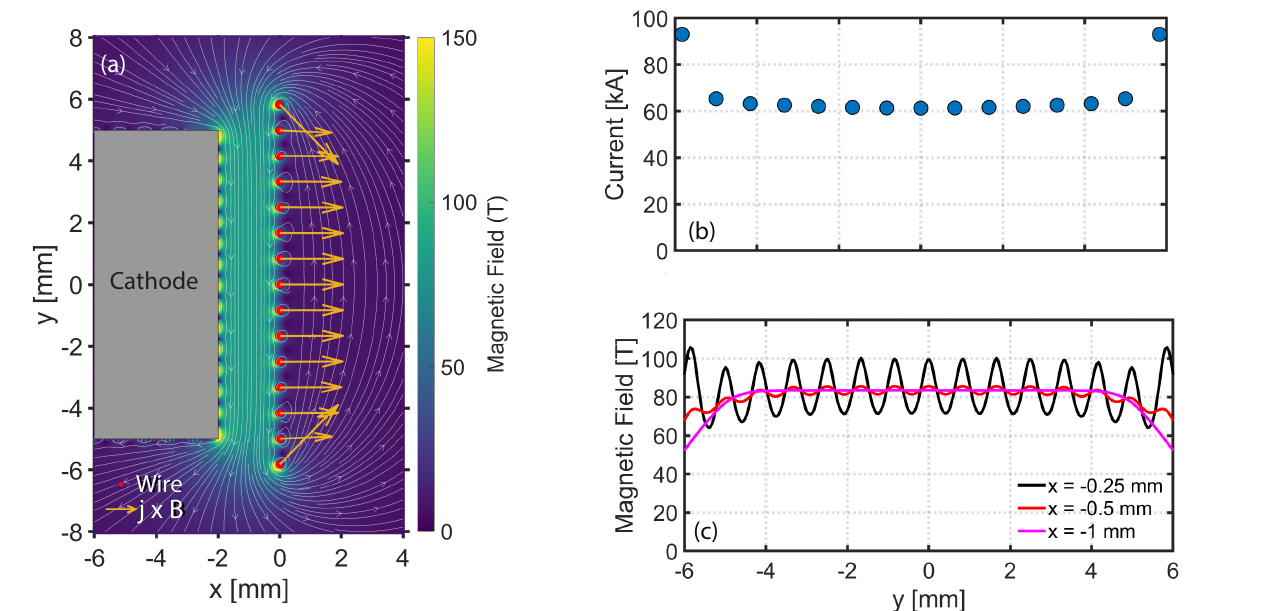}
\centering
\caption{
(a) Simulated magnetostatic magnetic field distribution in the load hardware. 
(b) Simulated current distribution in the wires at peak current, calculated from a magnetostatic inductance calculation. 
(c) Variation of the magnetic field strength in the AK gap at peak current at $x = -0.25, \, -0.5 \, \& \qty{-1}{\milli\meter}$.  
}
\label{fig:magnetostatic}
\end{figure*}

 \section{Experimental and Diagnostic Setup}
 \label{sec:setup}

\subsection{Load Hardware}

\label{sec:load}
 \autoref{fig:load} shows the load hardware configuration for this experiment. The load consists of a linear array of 15 equally-spaced aluminum wires. The wire-to-wire separation is $\qty{0.83}{\milli \meter}$, and the array height is $\qty{12}{\milli \meter}$. The wires are separated from a $\qty{10}{\milli \meter}$ wide stainless-steel cathode by a $\qty{2}{\milli \meter}$ wide AK gap, and are held in position by clamps on the anode plate and on the top of the cathode post. We perform a parametric study by varying the wire diameter between $\qty{33}{\micro \meter} \leq d_{\text{wire}} \leq \qty{100}{\micro \meter}$ for different experimental shots. The COBRA pulsed-power machine (Cornell University),\cite{greenly20081} when operated in long pulse mode, drives a 1 MA peak current pulse through the load.\cite{douglass2007structure,shelkovenko2007wire} A calibrated Rogowski coil placed around the central cathode monitors the current delivered to the load. \autoref{fig:current_pulse} shows the variation of the current delivered with time, as measured by the Rogowski coil. We show the current pulse averaged over 6 successive shots. The current pulse has a double-peaked structure, as it is formed by triggering two Marx generators with a delay between them. The first peak has a magnitude of about $0.75$ MA, and appears roughly 125 ns after current start, while the second peak has a magnitude of approximately $1$ MA, and appears 250 ns after current start.  The shot-to-shot deviation in the current pulse for this experimental series is $< 10\%$. 

To gain insight into the current distribution and driving magnetic field for the planar wire array, we perform magnetostatic inductance and Biot-Savart calculations of the load hardware.\cite{bland2004use} The magnetostatic magnetic field distribution in the planar wire array is shown in \autoref{fig:magnetostatic}a. The magnetic field inside the AK gap is nearly uniform with $y$-directed field lines, which curve around the outermost wires to form closed loops outside the array.  The mean driving magnetic field (at peak current) inside the AK gap is about $80-100$ T. \autoref{fig:magnetostatic}c shows lineouts of the magnetic field strength along the $y$-direction inside the AK gap. At the center of the gap ($x = \qty{-1}{\milli \meter}$), the driving magnetic field is uniform in the middle of the array  ($|y|\leq 4$ mm) with a strength of about $81$ T, but drops sharply to about $50$ T near the position of the outermost wires ($y = \pm \qty{6}{\milli \meter})$. This is because in contrast to previous experiments, \citep{bland2004use} we use a cathode whose width is smaller than the linear extent of the wires, which decreases the magnetic field strength around the outermost more inductively-favorable wires. Closer to the position of the wires, the magnetic field is dominated by the local magnetic field around each wire, resulting in a periodic variation in the field strength, as seen in \autoref{fig:magnetostatic}c.  Finally, unlike cylindrical exploding wire arrays, where field lines form closed loops inside the AK gap, and the field decays to zero outside the wires,\cite{lebedev2014formation} here the magnetic field lines must form closed loops outside the AK gap in the planar wire array. This means that there is a non-zero vacuum magnetic field in the flow region to the right of the wires, which is expected to be about $10\%$ of the driving magnetic field from the magnetostatic calculations.

The simulated current distribution in the wires at peak current (1 MA) is shown in \autoref{fig:magnetostatic}b. The current in the wires is symmetric about the $y = 0$ mm plane, and increases slightly with distance from the centerline for the inner wires. The current per wire is about $60-65$ kA for the inner wires, and increases sharply to approximately $90$ kA for the outermost wires. The higher current in the inductively-favorable outermost wires has also been reported previously in inductance and wire dynamics model computations of the planar wire array.\citep{bland2004use,esaulov2006wire} \citeauthor{bland2004use} considered both the resistive and inductive division of current between the wires, and found the experimental observations to be more consistent with the inductive current division, driving much higher current to the outermost wires.\citep{bland2004use} Due to the higher current, the rate of mass ablation from the outermost wires is expected to be higher. From a rocket model calculation,\cite{lebedev2002snowplow} assuming $\qty{50}{\micro \meter}$ diameter wires, we expect the outermost wires to ablate 50\% of their initial mass around $\qty{200}{\nano \second}$ after current start.


When current flows through the wires, the wires heat up resistively to generate a low-density coronal plasma surrounding the dense wire cores. The ablated plasma is accelerated by the ${\bf j \times B}$ force; this results in an outward flow of plasma into the region to the right of the wires. \autoref{fig:magnetostatic}a also shows the direction and relative magnitude of the ${\bf j \times B}$ force acting on the wire locations. The ${\bf j \times B}$ force at the wires points in the $+x$-direction for the inner wires, and its magnitude remains roughly consistent for the inner wires. The outer wires experience a ${\bf j \times B}$ force directed towards the center of the array. This is due to the bending of the field lines around the outer wires, as observed in \autoref{fig:magnetostatic}a. In designing the wire array, we explored different cathode sizes and AK gap widths in the magnetostatic simulations, however, the magnetic field always curves around the outer wires, similar to that in previous experiments,\citep{bland2004use} leading to an inward-directed a ${\bf j \times B}$ force. A shorter cathode relative to the linear extent of the wires, however, allows us to reduce the magnetic field strength around the higher current-carrying outermost wires, and thus, make the magnitude of the ${\bf j \times B}$ force relatively more uniform. 

\subsection{Diagnostic Setup}
\label{sec:diagnostic}

\begin{figure*}
\includegraphics[width=1.0\textwidth,page=3]{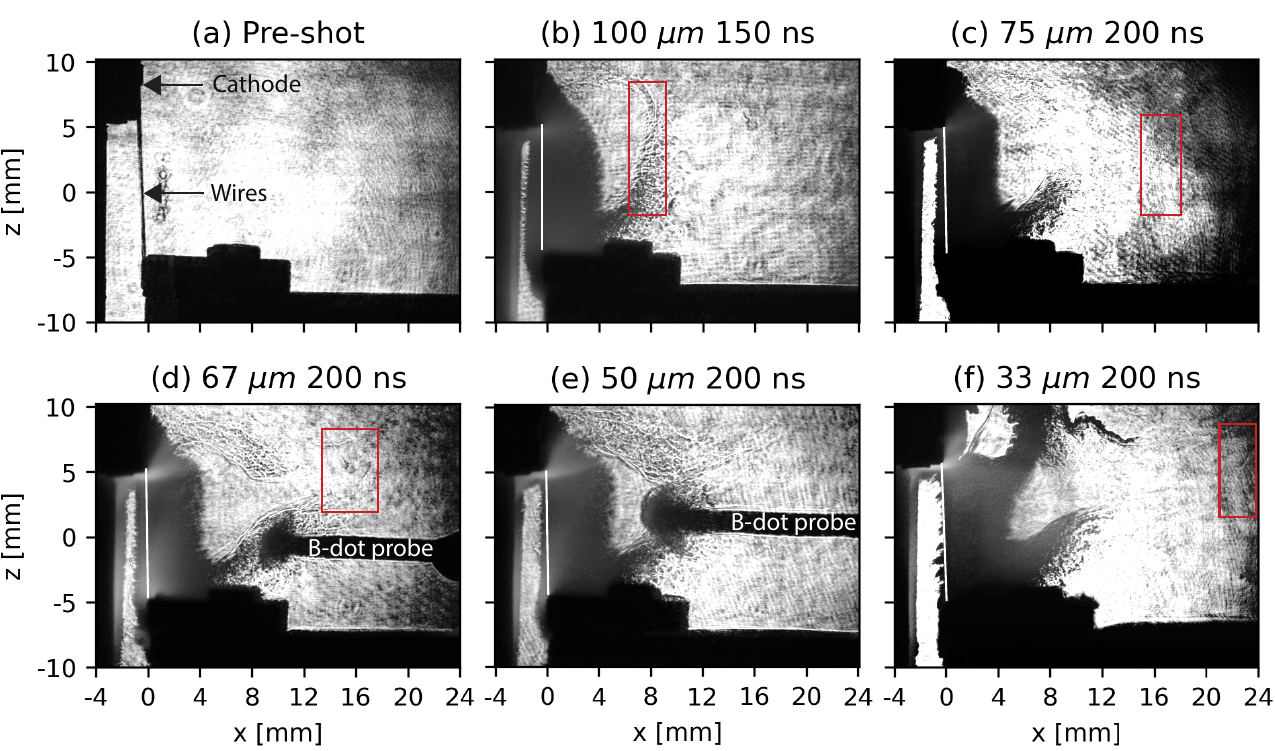}
\centering
\caption{
(a) Shadowgraph of the load hardware recorded before the start of the experiment. (b-f) Shadowgraphs of plasma ablation from the planar wire array for different wire diameters $\qty{33}{\micro \meter} \leq d_{\text{wire}} \leq \qty{100}{\micro \meter}$. In each image, plasma flow is from left to right. We indicate the initial position of the wires, determined from preshot images, with a white line. In (d) and (e), we also position b-dot probes in the flow.
}
\label{fig:shadowgraphs}
\end{figure*}

We use laser shadowgraphy to visualize the plasma flow from the planar wire array.  The shadowgraphy system is set up to provide a side-on view ($xz$ plane) of the experimental setup. This view is shown in \autoref{fig:shadowgraphs}a, which is a pre-shot image of the load. As the laser beam propagates through the plasma, electron density gradients deflect the light away from regions of higher density (lower refractive index) towards regions of lower density (higher refractive index). The intensity measured by the detector is thus related to gradients of electron density.\citep{Hutchinson2002}

In addition to shadowgraphy, we use a Mach-Zehder imaging interferometry system to measure the spatially-resolved line-integrated electron density of the plasma. Our interferometry system is set up to provide both an end-on ($xy$ plane) and a side-on view ($xz$ plane) of the experimental setup (see \autoref{fig:load}). When the probe beam propagates through the plasma, the resulting phase accumulated by the beam distorts the fringe pattern, and introduces a spatially-varying fringe shift, \citep{Swadling2013} which we use to reconstruct the phase difference between the probe and reference beams, and to determine the spatially-resolved line-integrated electron density. \citep{Hare2019} The field-of-view of our interferometer includes volume devoid of plasma, where the fringes remain undistorted. This region of zero fringe shift is chosen as the region of zero density. Both the shadowgraphy and interferometry systems use a $\qty{532}{\nano \meter}$ Nd:YAG laser ($\qty{150}{\pico \second}$ pulse width, $\qty{100}{\milli \joule}$) with a 1" diameter field-of-view. In the end-on system, the laser beam enters through a $\qty{26.4}{\milli \meter}$ diameter hole in the anode plate, as shown in \autoref{fig:load}c. The interferograms and the shadowgraphs are captured simultaneously using Canon EOS DIGITAL REBEL XS cameras. The interferometry and shadowgraphy systems record 1 frame per shot. The shots are reproducible, and we build up dynamics over multiple shots with identical initial conditions.

We also use a time-gated micro-channel plate (MCP) camera to capture extreme-ultraviolet (XUV) self-emission from the plasma. The camera captures 4 frames (\qty{10}{\nano \second} inter-frame time, \qty{5}{\nano \second} exposure time) recorded on isolated quadrants of the MCP via $\qty{200}{\micro \meter}$ diameter pinholes. The XUV camera looks onto the wires in the $yz-$plane, with an azimuthal viewing angle of $\qty{7.5}{\degree}$ with respect to the $x$-axis, and a $\qty{5}{\degree}$ polar angle to the horizontal ($xy$) plane. The diffraction-limited spatial resolution of the system, for photon energies between 10-100 eV, is about $\qty{180}{\micro \meter}-\qty{18}{\micro \meter}$, while the geometric resolution is about $\qty{300}{\micro \meter}$.

\begin{figure}
\includegraphics[width=0.48\textwidth,page=7]{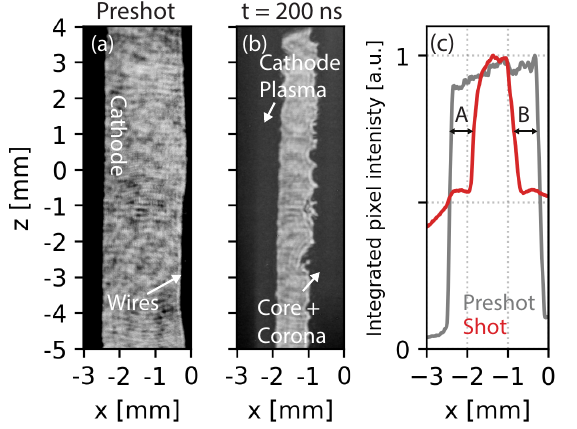}
\centering
\caption{
(a) Preshot shadowgraph of the AK gap for \qty{75}{\micro \meter} diameter wires. (b) Expansion of the cathode plasma, and the wire core and coronal plasma into the AK gap at 200 ns after current start. (c) Integrated pixel intensity from shadowgraphy images of the AK gap. The width labeled `A' represents the width of the cathode surface plasma, and `B' is the radius of the expanded coronal plasma.
}
\label{fig:AKgap1}
\end{figure}

\section{Results}
\label{sec:results}

\subsection{Shadowgraphs for different wire diameter}
\label{sec:shadow}

\autoref{fig:shadowgraphs} shows the side-on ($xz$-plane) shadowgraphs for different wire diameters $d_\text{wire} = 33-\qty{100}{\micro\meter}$. In each image, plasma flows from the left to the right, and we mark the initial position of the wires, determined from the pre-shot images, using a  white line. We record the shadowgraphs at 150 ns after current start for the \qty{100}{\micro \meter} diameter wire array, and at 200 ns for the $33-\qty{75}{\micro\meter}$ diameter arrays. In each shadowgraph, the wires expand to form an opaque region around the initial wire position. In this region, the propagating laser beam is lost, either because the density exceeds the critical density of the propagating light ($n_{e,crit} \approx \qty{4e21}{\per \centi \meter \cubed}$), or due to strong density gradients which refract the light out of the optical system's field of view. This dense region of plasma expands in the $+x$-direction, driven by the outwardly-directed ${\bf j \times B}$ force in the AK gap. Adjacent to the high-density region, we observe a region of relatively uniform density, followed by a narrow column of intensity fluctuations (indicated by a red rectangle in \autoref{fig:shadowgraphs}). This plasma column, consistent with observations of a magnetic precursor column in the literature,\citep{bland2004use,kantsyrev2008double} can be observed further away from the wires for the $33-\qty{75}{\micro\meter}$ diameter wires, compared to the \qty{100}{\micro \meter} wire array, which was recorded at an earlier time.

\begin{figure}
\includegraphics[width=0.48\textwidth,page=8]{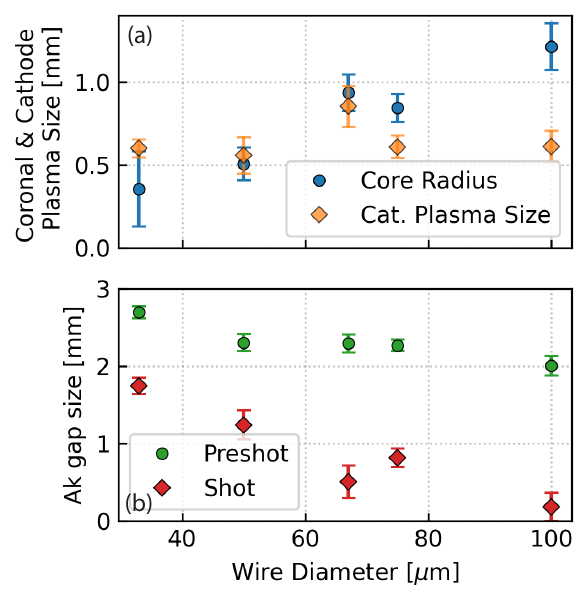}
\centering
\caption{
(a) Variation of the coronal radius and the cathode surface plasma width as a function of initial wire diameter. (b) Variation of the AK gap size with initial wire diameter. The range of values shown here comes from variation in the $z$-direction.
}
\label{fig:AKgap2}
\end{figure}

In addition to expansion in the $+x$-direction, the wire cores and the coronal plasma also expand into the AK gap. \autoref{fig:AKgap1} shows a magnified view of the AK gap for the $\qty{75}{\micro \meter}$ wire array, both before the experiment, and at 200 ns after current start. Reduction in size of the AK gap occurs due to wire core and coronal plasma expansion, as well as expansion of plasma from the cathode surface. The cathode surface plasma arises from current-driven ablation at the cathode surface, and photoionization via soft X-ray radiation generated by the wire cores. As observed in \autoref{fig:shadowgraphs}, the array with $\qty{100}{\micro \meter}$ wires exhibits the largest reduction in the AK gap size, while the AK gap remains  open for wire diameters $d_\text{wire} \leq \qty{75}{\micro \meter}$.

\begin{figure}
\includegraphics[width=0.48\textwidth,page=4]{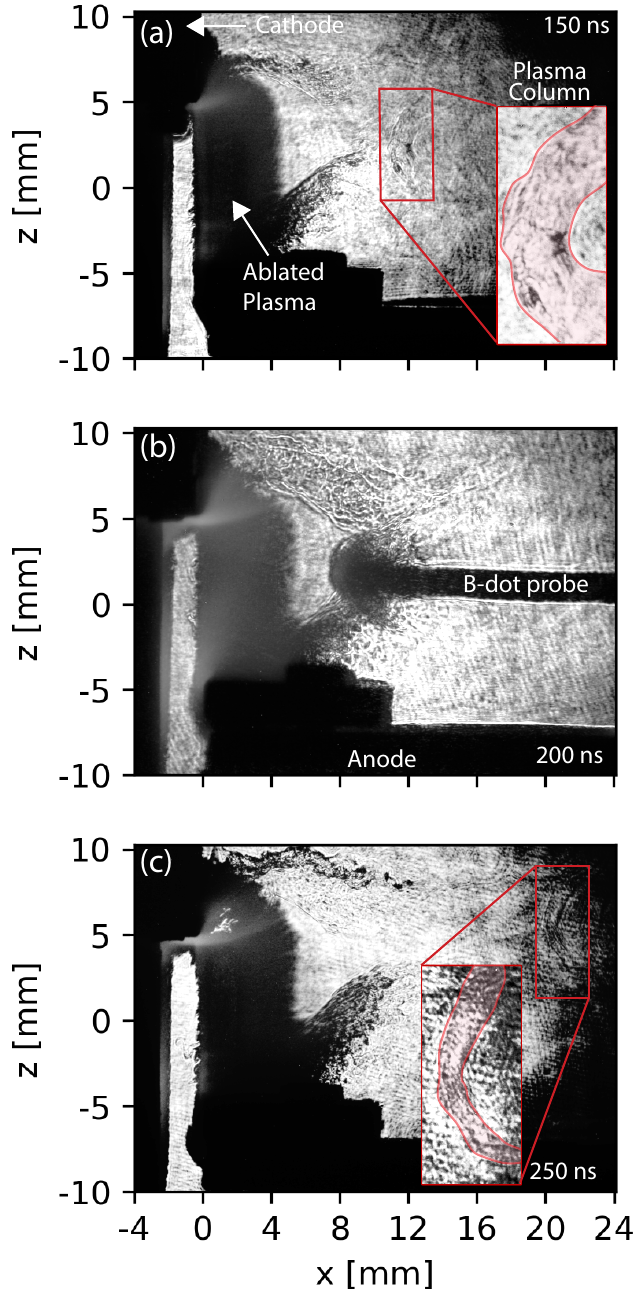}
\centering
\caption{
Plasma ablation from a planar wire array with $\qty{50}{\micro \meter}$ diameter wires at (a) 150 ns, (b) 200 ns, and (c) 250 ns after current start. Shadowgraphs are recorded in separate experimental shots. The red box shows the position of the plasma column, which travels at roughly $\qty{100}{\kilo \meter \per \second}$ between 150-250 ns. Note that in (b), we have placed a b-dot probe in the flow.
}
\label{fig:50um_shadow}
\end{figure}

We use the intensity of the shadowgraphs to estimate the diameter of the wire cores, and the size of the cathode surface plasma. We crop the shadowgraph to a smaller window which includes the cathode surface, the AK gap, and the expanding coronal plasma ($\qty{-3}{\milli \meter} \leq x \leq \qty{0}{\milli \meter}, \, |z| \leq \qty{5}{\milli \meter}$) (see \autoref{fig:AKgap1}b);  then integrate the pixel intensity along the $z$-direction. \autoref{fig:AKgap1}c shows the integrated pixel intensity as a function of position $x$ for the $\qty{75}{\micro \meter}$ diameter array, both for the preshot and the experimental shadowgraphs. The dark-to-light transitions in the preshot intensity profile represent the positions of the cathode surface and the wires, while those in the shot intensity profile represent the `edges' of the cathode plasma and the coronal plasma respectively. We fit a sigmoid function --- the cumulative density of a normal distribution --- to the light-to-dark intensity transitions, and determine the wire coronal and cathode plasma `edges' from the means $\mu$ of the fitted functions (or equivalently, the full-width-at-half-maximum of the transition). We estimate the uncertainty from the fitted function's standard deviation $\sigma$. We can then estimate the width of the cathode plasma from the distance between the cathode plasma edge and the position of the cathode surface (quantity A in \autoref{fig:AKgap1}c). Similarly, we determine the coronal plasma radius from the distance between the initial wire position and the position of the coronal plasma edge (quantity B in \autoref{fig:AKgap1}c).

\begin{figure}[b!]
\includegraphics[width=0.5\textwidth,page=15]{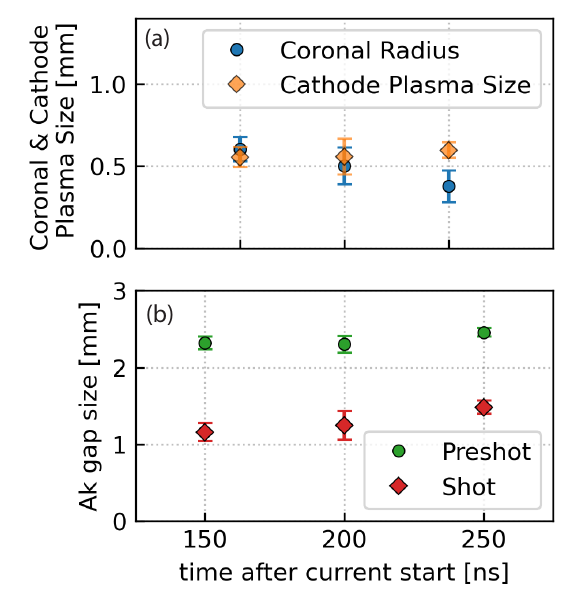}
\centering
\caption{
 (a) Temporal variation of the coronal radius and cathode plasma size for \qty{50}{\micro \meter} diameter wires. (b) Temporal variation of AK gap for \qty{50}{\micro \meter} diameter wires. Here, each data point comes from a separate experimental shot, and the range of values shown here comes from variation in the $z$-direction.
}
\label{fig:AKgap_vs_time}
\end{figure}

\begin{figure}
\includegraphics[width=0.48\textwidth,page=12]{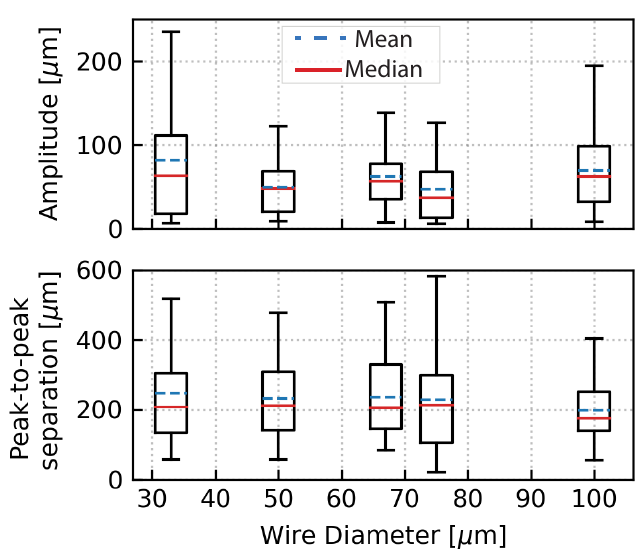}
\centering
\caption{
Variation of the mean amplitude and peak-to-peak separation of the axial instability in the wire core and coronal plasma as a function of initial wire diameter. Values as calculated at 200 ns after current start for wire diameters $33-\qty{75}{\micro \meter}$, and at 150 ns after current start for the $\qty{100}{\micro \meter}$ diameter wires. The red solid line and the blue dashed line represent the median and mean of the distribution respectively. The bottom and top sides of the rectangle represent the $25^{th}$ and $75^{th}$ percentile, while the end caps show the full range of the distribution.
}
\label{fig:instability}
\end{figure}

\autoref{fig:AKgap2}a shows the variation of the coronal plasma radius and the cathode plasma width with varying initial wire diameter. The coronal radius increases with increasing initial diameter of the wires, while the width of the cathode plasma remains relatively constant at roughly $\qty{0.5}{\milli \meter}$. \autoref{fig:AKgap2}b shows the variation of the width of the AK gap with the initial wire diameter. Consistent with the shadowgraphs in \autoref{fig:shadowgraphs}, the AK gap width decreases with increasing wire diameter. The $\qty{33}{\micro \meter}$ wires exhibit the smallest reduction in gap size, where the gap decreases from about $\qty{2.7}{\milli \meter}$ initially to roughly $\qty{1.8}{\milli \meter}$ during the experiment. In contrast, the $\qty{100}{\micro \meter}$ wires exhibit the largest decrease in gap size, from about $\qty{2}{\milli \meter}$ initially to roughly $\qty{0.2}{\milli \meter}$ 150 ns after current start. The smaller gap size for large wire diameters is primarily due to the increased core and coronal plasma radius of the larger wires. This is in contrast to \citeauthor{bland2004use}, in which the AK gap closure was almost entirely due to the expansion of plasma from the return electrode, and not from the thin \qty{7.5}{\micro \meter} diameter tungsten wires.

\begin{figure*}
\includegraphics[width=1.0\textwidth,page=5]{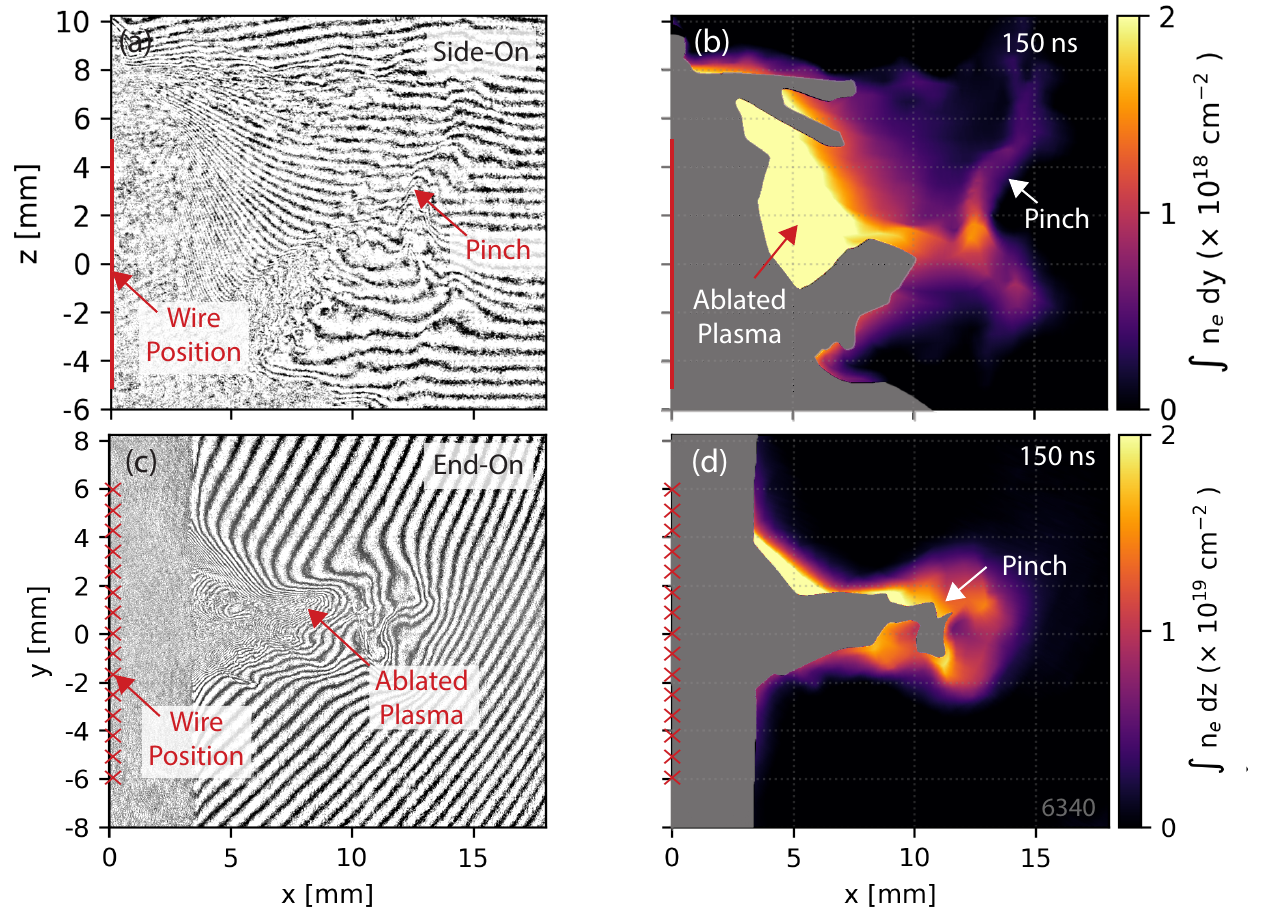}
\centering
\caption{
(a) Side-on raw interferogram for the \qty{50}{\micro \meter} diameter wire array at 150 ns, using a Mach-Zehnder interferometer with a 532 nm laser. 
(b) Side-on line-integrated electron density map determined from interferometry. 
(c) End-on raw interferogram at 150 ns after current start for the \qty{50}{\micro \meter} diameter wire array, recorded during the same experimental shot. 
(d) End-on line-integrated electron density map determined from interferometry. Regions in grey near the wires represent locations where the probing beam is lost. 
}
\label{fig:interferogram}
\end{figure*}

The methodology described above provides an upper limit on the wire core diameter. This is because the opaque region in the shadowgraph includes both the wire core and the surrounding coronal plasma. Moreover, density gradients in this region refract the light away from the relatively denser wire cores, resulting in a magnified image. Nevertheless, the shadowgraphs reflect the general trend observed in the variation of the core size with wire diameter.  X-ray backlighter imaging, which can probe deeper into the core region, can provide a better estimate of the core size. However, this diagnostic was not available for this experimental series. Previous experiments aimed at characterizing the core size report that values determined from shadowgraphs can be $5-10\times$ larger than that determined from simultaneous X-ray imaging. \cite{shelkovenko2007wire}

\subsection{Temporal evolution of ablation from the array}

We compare side-on shadowgraphs for $\qty{50}{\micro \meter}$ diameter wire arrays at 150 ns, 200 ns, and 250 ns in \autoref{fig:50um_shadow}. These shadowgraphs are recorded in separate experimental shots, with identical load hardware. The plasma column (red box in \autoref{fig:50um_shadow}) on the right of the image travels in the $+x-$direction, from $x \approx 12$ mm at 150 ns to $x \approx 22$ mm at 250 ns after current start. This corresponds to an average velocity of about $\qty{100}{\kilo \meter \per \second}$, which is consistent with the magnitude of flow velocity observed in previous wire array experiments. \citep{suttle2019interactions,datta2022time} The outward translation of the plasma column in our over-massed array is in contrast to that observed in the under-massed case, where the column remains mostly stationary ($V <\qty{15}{\kilo \meter \per \second}$) between the time of formation and implosion.\citep{bland2004use}  

The AK gap remains open throughout the experiment. The temporal evolution of the coronal plasma radius and the cathode plasma width are shown in \autoref{fig:AKgap_vs_time}a. The measured coronal radius decreases weakly with time, from about $ \qty{1.2}{\milli \meter}$ at 150 ns to about $\qty{0.75}{\milli \meter}$ at 250 ns after the current start. In contrast, the width of the cathode surface plasma remains roughly constant at about $\qty{0.5}{\milli \meter}$. Due to the decreasing coronal plasma radius, the AK gap also becomes slightly larger with time, as observed in \autoref{fig:AKgap_vs_time}b.  



\subsection{Instability Growth}

Axial perturbations of the coronal plasma appear in the AK gap, as observed in \autoref{fig:AKgap1}b. The presence of this axial instability is consistent with previous studies of wire array ablation. \citep{Lebedev1999,Chittenden2004}  In \autoref{fig:instability}, we characterize the amplitude and wavelength distribution of the instability as a function of the initial wire diameter. We determine the amplitude from half the peak-to-valley distance of the plasma-vacuum interface in the AK gap, which we characterize using the interface-detection technique similar to that described in Sec. \ref{sec:shadow}. Similarly, we estimate the wavelength of the instability from the peak-to-peak separation of the perturbations at the plasma-vacuum boundary.  In \autoref{fig:instability}, the red solid line and the blue dashed line represent the median and mean of the distribution respectively. The bottom and top sides of the rectangle represent the $25^{th}$ and $75^{th}$ percentile (i.e. the interquartile range), while the end caps show the full range of the distribution. The mean and median values for the perturbation amplitude are similar for most wire diameters, and remain largely invariant of the initial wire diameter, with a value of about $\qty{50}{\micro \meter}$. Both the $\qty{33}{\micro \meter}$ and $\qty{100}{\micro \meter}$ wires exhibit a relatively higher upper range of the perturbation amplitude, showing the existence of large amplitude perturbations. The wavelength distribution remains largely independent of the initial wire diameter, with a median peak-to-peak separation of about $\qty{200}{\micro \meter}$. We can also measure the temporal variation of the amplitude and wavelength from the shadowgraphs in \autoref{fig:50um_shadow}. Both the amplitude and wavelength of the instability exhibit little variation with time, which indicates saturation of the instability growth. We note that the shadowgraphs provide a line-integrated (along the  $y$-direction) view of the perturbation. Therefore, the process of extracting wavelength from the peak-to-peak separation, for the case where peaks from multiple wires overlap along the line-of-sight, becomes more complicated.

  \begin{figure*}
\includegraphics[width=1.0\textwidth,page=9]{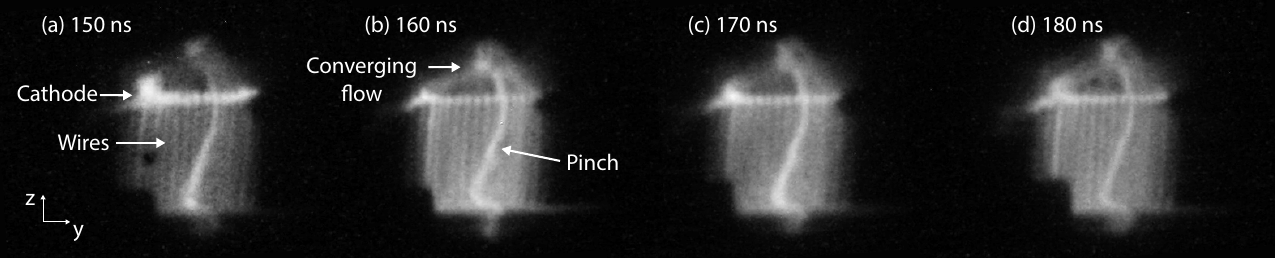}
\centering
\caption{
XUV self-emission images of a planar wire array with $\qty{50}{\micro \meter}$ diameter wires. The pinch appears as a column of bright emission.
}
\label{fig:XUV}
\end{figure*}

\subsection{Electron density measurements}

 \autoref{fig:interferogram}a \& b show the side-on ($xz$-plane) interferogram, together with the line-integrated electron density, recorded at $t = \qty{150}{\nano \second}$ after current start for the array with \qty{50}{\micro \meter} diameter wires. We indicate the initial position ($x = 0$ mm) of the wires using a red line. Close to the wires, the high-density plasma forms an opaque region where the probing beam is lost, similar to that in the side-on shadowgraphy images (\autoref{fig:shadowgraphs}). Adjacent to this opaque region, where the density is lower, interference between the probe and reference beams forms periodic bright and dark fringes. In this region, plasma flow from the wires distorts the fringe pattern, whereas, in regions devoid of plasma, the fringes appear undistorted. We trace the fringes by hand, and post-process the traced images in MAGIC2 to calculate the line-integrated electron density from the distortion of the fringes.\citep{Hare2019}  As expected due to time-of-flight effects, electron density is high close to the wires, and decreases with distance from the array.  At the plasma-vacuum boundary ($x \approx 12-15$ mm), the plasma forms a discontinuous column of enhanced electron density. The sharp rise in the electron density in this region indicates the presence of a shock-like structure. The width of the transition is about $\qty{2}{\milli \meter}$. The shape of the plasma column exhibits significant modulation in the axial direction, consistent with what we observe in the simultaneously-recorded shadowgraph of the load (\autoref{fig:50um_shadow}a). 

\autoref{fig:interferogram}c \& d show the end-on ($xy$-plane) interferogram and line-integrated density recorded 150 ns after current start. The probing laser beam in the region $x < 4$ mm is blocked by the load hardware; but the flow region $x \geq 4$ mm is illuminated via the laser feed shown in \autoref{fig:load}c. As seen in \autoref{fig:interferogram}a \& b, plasma flow emanating from the wires is redirected towards the center ($y = 0$ mm) of the array. The converging flows collide or `pinch', forming a region of enhanced density, roughly 12-\qty{15}{\milli \meter} from the wires. The pinch has also been typically referred to as the `magnetic precursor column' in wire array literature, as it is the precursor to the final implosion phase which we suppress in these experiments by over-massing the wire array. \cite{bland2004use} The position of the pinch is consistent with that of the column of enhanced electron density observed in the side-on electron density map (\autoref{fig:interferogram}b). 


In  \autoref{fig:interferogram}b, at $x \approx 4$ mm from the wires, the line-integrated density is $\langle n_e L_y \rangle \approx 5-\qty{6e18}{\per \centi \meter \squared}$, which falls to $\langle n_e L_y \rangle \approx \qty{0.4e18}{\per \centi \meter \squared}$ at 11 mm from the wires, right before the position of the pinch. From end-on interferometry (\autoref{fig:interferogram}d), we estimate the integration length scale $L_y$ by computing the extent of the plasma in the $y$-direction. This gives us values of $L_y (x = \qty{4}{\milli \meter}) \approx \qty{8}{\milli \meter}$, and $L_y (x = \qty{11}{\milli \meter}) \approx \qty{4}{\milli \meter}$. The average electron densities, inferred from $\langle n_e L_y\rangle / L_y$, are therefore $\bar{n}_e  \approx \qty{4e18}{\per \centi \meter \cubed}$ at $x = \qty{4}{\milli \meter}$, and $\bar{n}_e \approx \qty{1e18}{\per \centi \meter \cubed}$ at $x = \qty{11}{\milli \meter}$ from the wires. In  \autoref{fig:interferogram}b, the pinch exhibits a line-integrated density of $\langle n_e L_y \rangle \approx  \qty{0.8e18}{\per \centi \meter \squared}$ at approximately \qty{13}{\milli \meter} from the wires. Assuming a length scale $L_y \approx \qty{4}{\milli \meter} $, the average electron density in the pinch is $\bar{n}_e \approx \qty{2e18}{\per \centi \meter \cubed}$. This represents a roughly $ 2\times$ jump in the electron density at the pinch compared to the flow upstream of the pinch.

 \begin{figure}[b!]
\includegraphics[width=0.48\textwidth,page=14]{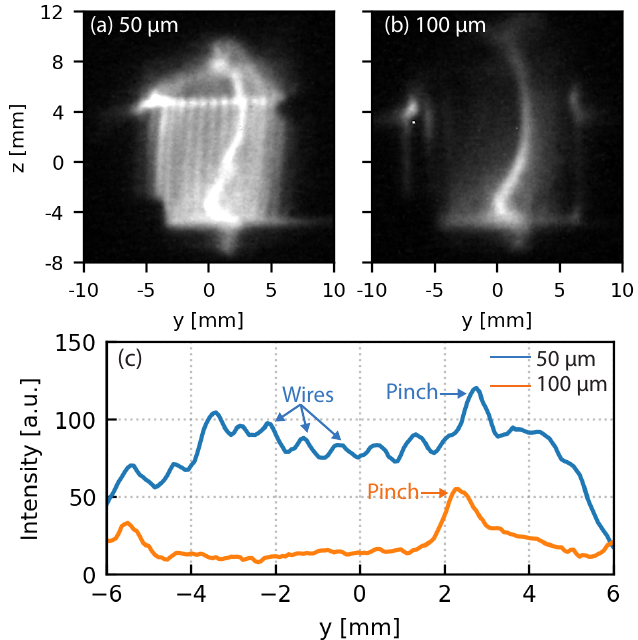}
\centering
\caption{
XUV self-emission images of a planar wire array with  (a) $\qty{50}{\micro \meter}$ diameter wires, and (b) $\qty{100}{\micro \meter}$ diameter wires. Wires are easily distinguishable in the $\qty{50}{\micro \meter}$ case, but not for the $\qty{100}{\micro \meter}$ diameter wires. (c) Lineouts of intensity along $z = 4$ mm for $\qty{50}{\micro \meter}$ and  $\qty{100}{\micro \meter}$ diameter wires.
}
\label{fig:XUV_2}
\end{figure}

\subsection{XUV Self-Emission}

XUV self-emission images from the load hardware with the $\qty{50}{\micro \meter}$ diameter wires are shown in \autoref{fig:XUV}. The wires and the cathode appear as regions of bright emission. Emission from the inner wires appears uniform in intensity, indicating roughly equal current distribution in the wires, as predicted by the magnetostatic calculation (\autoref{fig:magnetostatic}). The outer wires, however, appear dimmer, which may indicate that the current has switched to the inner wires due to the higher initial rate of ablation from the outer wires, as indicated by the rocket model calculation in Sec. \ref{sec:load}. Although the wires appear as well-separated columns of emission, the resolution of the optical system prevents us from making quantitative measurements of the core diameter from the XUV images. We observe that the flows from the wires converge to form the pinch, which appears as a brightly-glowing column oriented in the $z$-direction. The increased emission from the pinch is consistent with its higher electron density (Figure \ref{fig:interferogram}). Furthermore, shock heating and Ohmic dissipation in the pinch may also contribute to a higher temperature, and consequently, higher radiative emission. The XUV images also exhibit the axial non-uniformity in the shape of the pinch, consistent with the interferometry and shadowgraphy results (\autoref{fig:50um_shadow}a and \autoref{fig:interferogram}b). Finally, the structure of the plasma ablation and the pinch remains roughly invariant across the different frames over the observation window of $150-\qty{185}{\nano \second}$. This is in contrast to the shadowgraphy and interferometry images (\autoref{fig:50um_shadow} \& \autoref{fig:interferogram}) which show significant ($V \approx \qty{100}{\kilo \meter \per \second}$) motion of the pinch. This may indicate that this is a `ghost' image of the pinch, recorded when the MCP is not triggered, due to radiation bleed-through at the time when emission from the pinch is at a maximum.

In \autoref{fig:XUV_2}, we compare the XUV emission from the planar wire arrays with  $\qty{50}{\micro \meter}$ and  $\qty{100}{\micro \meter}$ diameter wires respectively. In both cases, the pinch is visible as a bright column of emission, and the shape of the pinch is similar between the two images. For the $\qty{50}{\micro \meter}$ diameter case, the wires appear as discrete columns of enhanced emission, as can be observed in lineouts of the intensity at $z = 4$ mm (\autoref{fig:XUV_2}c). In contrast, the wires are not easily distinguishable for the thicker $\qty{100}{\micro \meter}$, consistent with the core size becoming comparable to the inter-wire separation, as seen in \autoref{fig:AKgap2}. 


\section{Discussion of results}
\label{sec:discussion}

\subsection{AK Gap and Wire core size}

Previous wire array experiments with $\sim\qty{10}{\micro \meter}$ diameter wires show that the diameters of the wire cores and the surrounding corona both increase with initial wire diameter, and are largely independent of the current per wire and the inter-wire separation. \cite{shelkovenko2007wire} Our experimental results are consistent with this effect --- \autoref{fig:AKgap2}a exhibits a roughly linear increase in the coronal radius with increasing wire diameter, and the measured coronal diameter is roughly $20-25\times$ the initial wire diameter. While the coronal radius increases with the initial wire diameter,  the size of the cathode plasma remains relatively constant (see \autoref{fig:AKgap2}a). This is expected since changing the initial wire diameter is not likely to affect the current distribution through the cathode. The larger coronal radius is, therefore, the primary reason for gap closure in the thick $\qty{100}{\micro \meter}$ case. The gap closes at 150 ns after current start, which makes $d_\text{wire} = \qty{100}{\micro \meter}$ an undesirable wire diameter, both in these planar wire experiments and on the Z experiments for which these experiments are a scaled test. Furthermore, the coronal radius also becomes larger than the inter-wire separation in this case, which inhibits plasma ablation and magnetic field advection from the array.\cite{velikovich2002perfectly} 

The early gap closure for the $\qty{100}{\micro \meter}$ diameter case could be a consequence of lower Ohmic heating in the skin region around the wire core. The initial electrical explosion of the wires forms a dense cold wire core consisting of vapor and microscopic liquid metal droplets. Without further Ohmic heating by the current, we would expect the wire core radius $R(t)$ to expand isotropically at a rate comparable to its local sound speed $C_{\text{core}}$ into the vacuum, i.e. $dR/dt \sim C_{\text{core}}$. For thin wires, the current flowing over the wire core surface in the skin region Ohmically heats the material at the edge of the core, forming coronal plasma that is redirected by the global $\mathbf{j \times B}$ force. However, if the wire core is sufficiently large,  current density in the skin region $j_{\text{skin}} = I(t)/(2 \pi R(t) \delta)$ will be lower. Here, $I(t)$ is the driving current, and $\delta = \sqrt{2 \eta/\omega \mu}$ is the resistive skin depth, which depends on the material resistivity $\eta$, the angular frequency $\omega$ of the driving current, and the medium permeability $\mu$. Consequently, for a large initial wire diameter, the Ohmic heating rate $\eta j_{\text{skin}}^2$ may be too small to ionize all of the expanding gas. This will allow neutral gas expanding out of the wire core to remain unionized, and thus, unaffected by the $\mathbf{j \times B}$ force expelling it from the AK gap. It is this neutral gas which may be responsible for the observed gap closure. In future experiments, the importance of neutral gas expansion could be tested by exploiting the different refractive indices of plasma and neutral gas using two-color optical measurements.\citep{muraoka2016laser}

For the \qty{50}{\micro \meter} diameter wire array, the AK remains open late in time (t = 250 ns), and the coronal radius does not exceed the inter-wire separation, which is desirable for good ablation from the array (\autoref{fig:AKgap_vs_time}). In imploding wire arrays, the coronal radius increases initially in time, and then saturates to a constant value. \cite{shelkovenko2007wire} The time of saturation, typically $80-100$ ns after current start, corresponds to a change in the magnetic field topology around the wires, when the driving global ${\bf j \times B}$ force becomes strong enough to overcome the expansion of the coronal plasma, and redirects it to generate ablation streams. \cite{shelkovenko2007wire} In our experiments, we image the wires after the expected time of saturation, and therefore, do not expect significant temporal variation in the size of the wire cores between 150-250 ns after current start. As observed in \autoref{fig:AKgap_vs_time}a, the coronal plasma radius decreases weakly with time at a rate of about $\qty{2}{\micro \meter \per \nano \second}$ for \qty{50}{\micro \meter} diameter wires. 


The axial instability of the wires is ubiquitous in wire array z-pinch experiments, and is thought to be a modified $m=0$-like instability, which exhibits a constant amplitude and wavelength later in time, and is largely independent of initial wire diameter, and current per wire.\cite{lebedev2002snowplow,chittenden2004equilibrium, jones2005study} The time of saturation of the instability also corresponds to the time at which the wire cores cease to grow.\cite{chittenden2004equilibrium, jones2005study} In \autoref{fig:instability}, we observe that the distributions of the amplitude and the peak-to-peak separation of the perturbations remain largely independent of the initial wire diameter, consistent with observations of the axial instability in imploding wire array z-pinches. The amplitude and the peak-to-peak separation also exhibit minimal variation in time between 150-250 ns, which is well after the expected time of saturation of the instability ($\sim 80-100$ ns).\cite{jones2005study,shelkovenko2007wire}


 \begin{figure}
\includegraphics[width=0.48\textwidth,page=13]{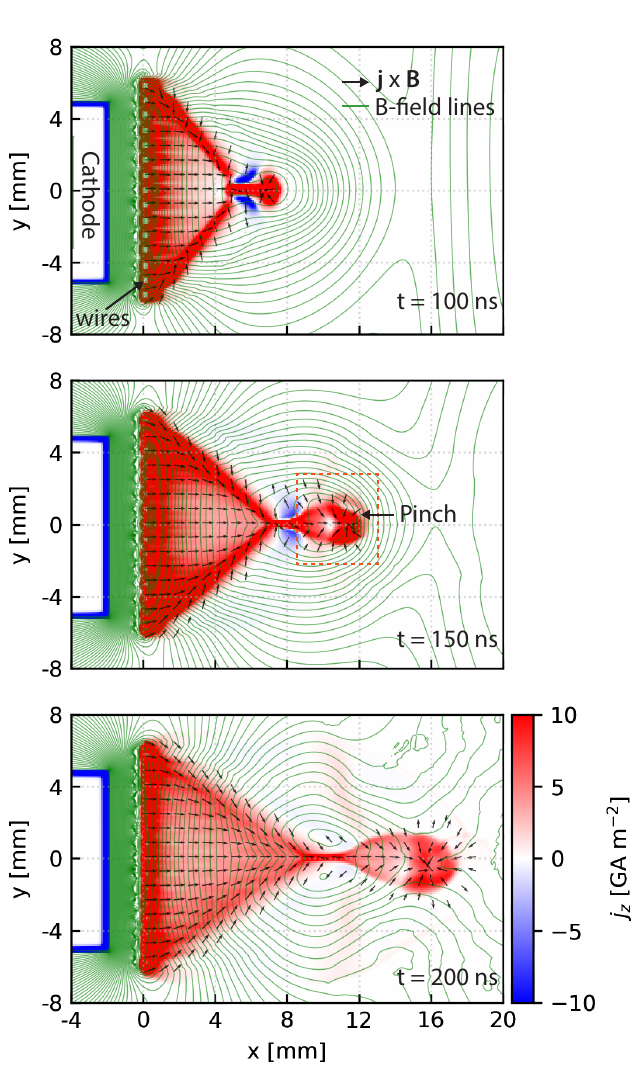}
\centering
\caption{
Current distribution in the load hardware from 3D resistive MHD simulations of the experiment. Here, we show the current distribution on a slice through the array midplane ($z = \qty{0}{\milli \meter}$). The black arrows represent the direction of the ${\bf j \times B}$ force, while the green lines are contours of the $z$-component of the magnetic vector potential $A_z$, which we use to represent the magnetic field lines.
}
\label{fig:sim_current}
\end{figure}

\begin{figure*}
\includegraphics[width=1.0\textwidth,page=6]{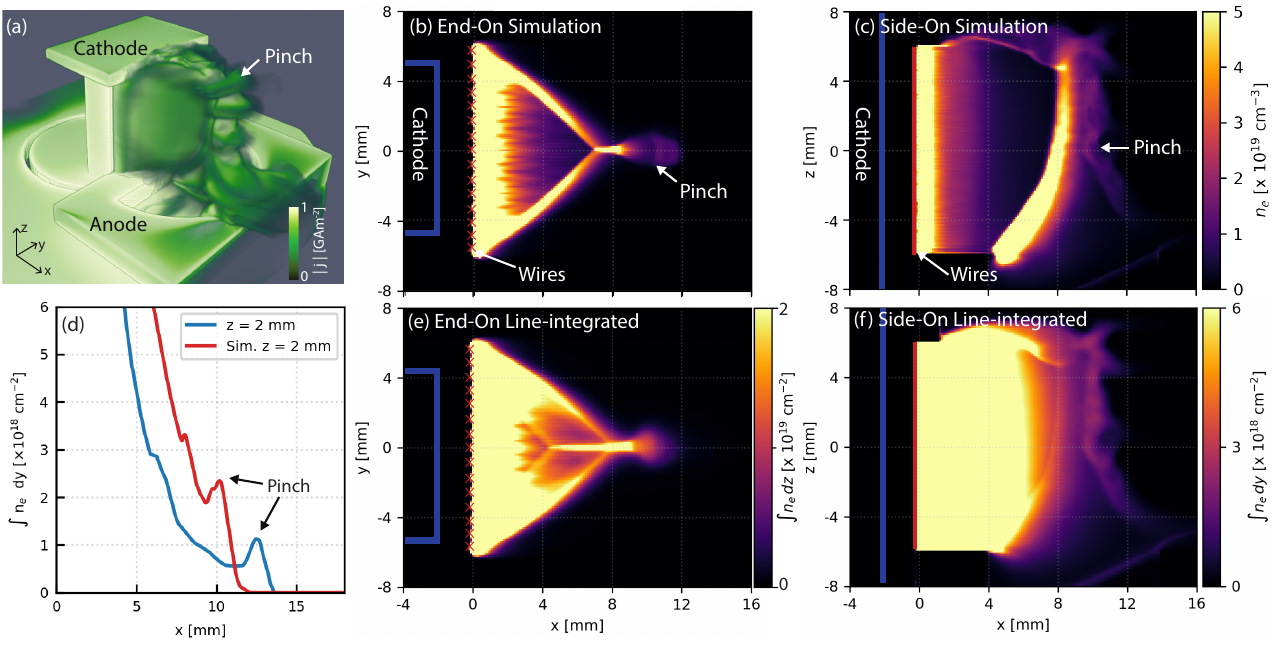}
\centering
\caption{
(a) Simulated current distribution in the load hardware at \qty{150}{\nano \second} after the current start. (b) End-on ($xy$-plane) slice of the simulated electron density at the array mid-plane at 150 ns. (c)  Side-on ($xz$-plane) slice of the simulated electron density at the array mid-plane at 150 ns. (d) Comparison of the line-integrated (along $y$) electron denisty between teh experiment and the simulation. (e) End-on line-integrated electron density. (f) Side-on-line integrated electron density. Plasma flow is from left to right. Wire positions are indicated by X's in (a) \& (c), and by red lines in (b) \& (d). The simulation shows converging flows, and the formation of a pinch roughly \qty{8}{\milli \meter} from the wires.
}
\label{fig:sim_ne}
\end{figure*}

\subsection{Pinch formation and comparison with 3D resistive MHD simulations}
\label{sec:sim}

To obtain greater insight into the ablation process, we perform three-dimensional simulations of the planar wire array using the resistive magnetohydrodynamic (MHD) code GORGON, a 3D (cartesian, cylindrical, or polar coordinate) Eulerian resistive MHD code with van Leer advection, and separate energy equations for ions and electrons. \cite{Chittenden2004, ciardi2007evolution} We use an optically thin recombination-bremsstrahlung radiation loss model, modified with a constant multiplier to account for line radiation, and a Thomas-Fermi equation-of-state to determine the ionization level.\cite{ciardi2007evolution} We simulate a planar wire array with the same geometric dimensions and wire material as in the experimental setup. The current pulse applied to the load was determined from a three-term sum-of-sines fit [$\sum_i a_i \sin(b_i t + c_i)$] to the integrated Rogowski signal shown in \autoref{fig:current_pulse}.  The simulation domain is a cuboid with dimensions $51.2 \times 50.4 \times 38$ \qty{}{\milli\metre\cubed}. We use an initial wire diameter of $\qty{50}{\micro \meter}$ in the simulation, with the initial mass of the wire distributed over a $\qty{400}{\micro \meter}$ diameter circular pre-expanded wire core. The simulations are performed with a grid size of $ \qty{50} {\micro \metre}$.  The driving magnetic field, calculated from Ampere's law, is applied as a boundary condition at the bottom-most cells in the simulation domain between the anode and cathode of a coaxial transmission line. The load geometry is implemented as stationary realistic conductivity electrode material on top of the coaxial line. 

\autoref{fig:sim_current} shows the simulated current density distribution at a slice through the array mid-plane ($z = 0$ mm) at 100, 150, and 200 ns after current start. As expected due to the skin effect, current density is concentrated on the outer surfaces of the cathode and the wires. The plasma from the outermost wires carries significantly more current than that from the inner wires, consistent with our magnetosonic prediction (\autoref{fig:magnetostatic}). Similar to the experiment, the converging plasma flows collide at $y = \qty{0}{\milli \meter}$ mid-plane to form the pinch. The pinch appears as a region of high current density, comparable to that in the wires. \autoref{fig:sim_current}a shows a three-dimensional rendering of the current distribution in the load at $t = \qty{150}{\nano \second}$. The pinch carries a significant amount of current, and provides a secondary path for the current to close between the anode and the cathode. In our simulation, the current in the pinch, determined from the area integral of the current in the dashed box shown in Figure \ref{fig:sim_current}, is roughly $ 30\%$ of that in the wires. This is consistent with the estimate of $30-40\%$ provided by \citeauthor{bland2004use}\citep{bland2004use}

\autoref{fig:sim_current} also shows the distribution of magnetic field lines in the planar wire array. We determine the field lines from contours of the $z$-component of the magnetic vector potential $A_z$. Near the AK gap, the magnetic field topology is similar to that calculated from our magnetostatic simulation (see \autoref{fig:magnetostatic}). The magnetic field lines are straight and uniform inside the AK gap, and bend around the outer wires to form closed field lines outside the array. The field inside the AK gap is significantly stronger than that outside, as observed from the relatively high density of lines in this region. The ablating plasma advects some of the magnetic field from inside of the array to the outside. Near the inner wires, the advected magnetic field lines are straight and uniform, oriented along the $y$-direction. However, away from the centerline ($y = \qty{0}{\milli \meter}$) and toward the edges of the plasma flow, the field lines bend, driven by the inward-directed ablation from the outermost wires. The spatial variation of the driving magnetic field along the $z$-direction inside the AK gap is small ($<5\%$).  

The arrows in \autoref{fig:sim_current} show the direction of the ${\bf j \times B}$ force acting on the ablated plasma. Near the inner wires, where the magnetic field is oriented in the $y$-direction, the force is directed in the $+x$-direction, whereas at the outermost wires, the curvature of the magnetic field results in a force directed towards the center of the array. As the plasma propagates away from the wires, the bending of the field lines due to the flows from the outermost wires results in an inward-directed (along the $x$-direction) ${\bf j \times B}$ force. This drives the collision of the plasma flows emanating from the wires, and the formation of the pinch. Similar to a traditional z-pinch, the magnetic field lines form closed circular loops, and the ${\bf j \times B}$ force is directed towards the center of the pinch. However, unlike a traditional z-pinch, the pinch experiences both mass and momentum injection from the left side of the pinch. Driven by the magnetic and thermal pressure of the plasma behind the pinch and the magnetic tension of the bent field lines, the pinch accelerates in the $+x$-direction. In the simulation, the center of the pinch travels about $\qty{8}{\milli \meter}$ between $150-\qty{250}{\nano \second}$, resulting in an average velocity of $\qty{80}{\kilo \meter \per \second}$. This is about $20\%$ lower than that inferred from the translation of the pinch in the experimental shadowgraphs (\autoref{fig:50um_shadow}). The flow upstream of the pinch is both supersonic ($M_S \approx 6$) and super-Alfvénic ($M_A \approx 2$), similar to that observed in previous pulsed-power-driven experiments of aluminum wire arrays.\cite{burdiak2017structure,russell2022perpendicular,datta2022structure} Due to the high current density in the pinch, it is a site of strong Ohmic dissipation. The electron temperature inside the pinch is $T_e \approx 100$ eV, which is significantly higher than that in the plasma flow behind the pinch $T_e \approx 6.5$ eV. The temperature of plasma near the wires is also about 5 eV, which may explain why the wires appear dimmer in the XUV images compared to the pinch (\autoref{fig:XUV}).

\autoref{fig:sim_ne}b shows a slice of the simulated electron density at the array mid-plane ($z = \qty{0}{\milli \meter}$) at 150 ns after current start. The electron density distribution appears similar to that in the experiment (\autoref{fig:interferogram}). Electron density is high closer to the wires, and falls with increasing distance in the $x$-direction, consistent with time-of-flight effects. The plasma flow from the inner wires is directed outwards along the $x$-direction, while that from the outermost wires is directed towards the center of the array. The flow converges, similar to that in the experiment, to form a pinch. \autoref{fig:sim_ne}c shows a side-on slice of the electron density through the array mid-plane ($y = \qty{0}{\milli \meter}$). The pinch appears as a discontinuous region of enhanced electron density at the vacuum-plasma boundary, similar to that in the experiment. Immediately behind the pinch, both the end-on and side-on slices show a region of lower density, consistent with the relatively-uniform intensity region observed in the experimental shadowgraphs (\autoref{fig:50um_shadow}). The electron density inside the pinch is $n_e \approx \qty{4e19}{\per \centi \meter \cubed}$, while that in the plasma flow just behind the pinch is $\qty{2e18}{\per \centi \meter \cubed}$. Our experimentally inferred value of the density behind the pinch is consistent with the simulation, while that for the pinch is lower than the density observed in the simulation. This may indicate that the integration length scale used to determine the experimental estimate of density is an overestimation of the true value. As observed in \autoref{fig:sim_ne}a, the width of the simulated pinch is approximately $\qty{1.5}{\milli \meter}$, whereas we estimate a value of about $\qty{4}{\milli \meter}$ from our experimental end-on density map (\autoref{fig:interferogram}d). This discrepancy can result from line integrating through the axially-modulated pinch, which widens the observed width of the pinch in the line-integrated density map (\autoref{fig:interferogram}). Using an integration length of \qty{1.5}{\milli\meter} results in an electron density of $\bar{n}_e \approx \qty{5e18}{\per \centi \meter \cubed}$, which is closer to, although still lower than, the density of the pinch in the simulation.

\autoref{fig:sim_ne}e and \autoref{fig:sim_ne}f show the line-integrated electron density in the end-on ($xy$-plane) and side-on ($xz$-plane) planes respectively. In \autoref{fig:sim_ne}d, we compare line-outs of the simulated electron density integrated along the $y$-direction with that from the experiment at $z = \qty{2}{\milli \meter}$. In both the experiment and the simulation, the line-integrated density falls with distance from the wires, and the pinch appears as a local enhancement of the density at the plasma-vacuum boundary. The magnitude of the line-integrated electron density in the simulation is comparable to that from the experiment. Note that the sharp increase in density at the pinch, as observed in \autoref{fig:sim_ne}c, is muted by line integration. The high density and the temperature of the pinch both contribute to the strong emission from the pinch, visible in the experimental XUV self-emission images (\autoref{fig:XUV}). 

In contrast to the experiment, where the pinch is located at $x \approx 12-15$ mm from the wires, the simulated pinch is closer to the wires ($x \approx 10$ mm) at this time. The slower velocity of the pinch in the simulation indicates a comparatively smaller driving force behind the pinch. We expect the pinch to be driven outwards due to the magnetic and thermal pressures of the plasma behind it, and by the magnetic tension of the bent field lines. A simple similarity argument can be used to show that the characteristic velocity of the pinch should be comparable to the local magnetosonic velocity $V_{pinch}^2 \sim (V_A^2 + C_S^2)$. Here, $V_A$ is the Alfvén speed, and $C_S$ is the sound speed of the plasma right behind the pinch. Previous comparison of experimental results with simulations indicates that the local thermodynamic equilibrium Thomas-Fermi model implemented in the simulation underestimates the temperature of the plasma.\cite{burdiak2017structure} The Alfvén speed $V_A = B/\sqrt{\mu_0 \rho}$ is a function of the magnetic field, which in turn, depends on the magnetic Reynolds number $R_m = UL/\bar{\eta}$. Here, $U$ and $L$ are the characteristic velocity and length scales of the plasma, $\rho$ is the mass density, and $\bar{\eta} \sim \bar{Z}T_e^{-3/2}$ is the magnetic diffusivity, which varies with the electron temperature $T_e$ and the average ionization $\bar{Z}$ of the plasma. A lower temperature leads to a lower $R_m$, and thus a relatively smaller advected field. This can reduce the magnetic pressure behind the pinch, and therefore contribute to a smaller velocity. In future experiments, optical Thompson scattering could be used to simultaneously characterize the velocity and temperature of the plasma. \cite{suttle2021collective}

Finally, in both the experiment and the simulation, the shape of the pinch exhibits an axial non-uniformity. As observed in \autoref{fig:interferogram}b and \autoref{fig:sim_ne}c, the pinch appears closer to the wires at the bottom of the array, and further away near the top. Our simulations indicate that this axial non-uniformity in the shape of the pinch may be related to flow over the extended anode plate (see \autoref{fig:load}), which prevents the expansion of the pinch at the bottom of the array, and also modifies the current distribution in the electrodes. When the simulations are repeated without an extended anode plate, the axial modulation in the pinch structure is reduced. In future experiments, we can mitigate this effect by exploring array geometries that do not require the extended anode plate.


\section{Conclusions}

In this paper, we explore the use of an over-massed planar wire array as a platform for laboratory astrophysics experiments, and as a scaled experiment to investigate the ablation of thick wires in cylindrical wire arrays driven by 10 MA current pulses. We characterize the ablation of plasma from a planar wire array fielded on the COBRA pulsed-power machine (1 MA, 250 ns rise time). The wire array comprises a linear arrangement of 15 equally-spaced aluminum wires separated from a planar cathode surface by a 2 mm AK gap. The planar wire array is designed to provide a driving magnetic field ($80-100$ T) and current per wire distribution (about $60-65$ kA), similar to that in a $\sim 10$ MA cylindrical exploding wire array fielded on the Z pulsed-power machine. Magnetostatic calculations show that the driving magnetic pressure inside the AK gap at peak current ($1$ MA) is about $81$ T, which is higher than that in a typical cylindrical wire array fielded on 1-MA university scale facilities (about $20-40$ T). In contrast to previous planar wire array experiments, the wire arrays are over-massed, so that they provide continuous ablation for the duration of the experiment, without experiencing the implosion stage.

We perform a parametric study by varying the initial wire diameter between $33-\qty{100}{\micro \meter}$. Laser shadowgraphy images show that the largest wire diameter (\qty{100}{\micro \meter}) exhibits early closure of the AK gap (150 ns after current start), while the gap remains open during the duration of the experiment for wire diameters between $33-\qty{75}{\micro \meter}$. The early closure of the AK gap for the \qty{100}{\micro \meter} diameter case is primarily due to the larger coronal radius of the wires, which may be a consequence of reduced Ohmic heating in the skin region surrounding the wire cores. For these large diameter wires, the coronal radius also becomes comparable to the AK gap size and the inter-wire separation, which is undesirable for good ablation from the wire array. Axial instabilities appear in the vacuum-plasma interface in the AK gap. The distributions of the amplitude and peak-to-peak separation of the perturbations remain largely invariant of the initial wire diameter, as has been previously observed on imploding and exploding cylindrical wire arrays.

Laser interferometry and time-gated XUV imaging are used to probe the plasma flows. Plasma ablating from the wires is redirected towards the array mid-plane ($y = 0$ mm), and the resulting collision of the converging flows generates a pinch, which propagates away from the wires at an average velocity of about $\qty{100}{ \kilo \meter \per \second}$. The pinch appears as a discontinuous column of enhanced plasma density ($\bar{n}_e \approx \qty{2e18}{\per \centi \meter \cubed}$) and strong XUV emission. Three-dimensional resistive MHD simulations reproduce the primary characteristics of the ablation observed from the experiments. Visualization of the current density and magnetic field in the load demonstrates that flows converge under the action of a pinching ${\bf j \times B}$ force. This arises from the bending of magnetic field lines due to the inward-directed flows from the outermost wires. The pinch is a site of high current density, and exhibits a magnetic field topology similar to that of a z-pinch. The simulated pinch also exhibits a significantly higher temperature, compared to the plasma behind it, which combined with the enhanced density, accounts for the strong XUV emission observed in the experiment.

\section{Acknowledgements}
The authors would like to thank Todd Blanchard and Harry Wilhelm for their work in support of the experiments. This work was funded by NSF and NNSA under grant no. PHY2108050, and by the EAGER grant no. PHY2213898. Simulations were performed on the Engaging cluster funded by DE-FG02-91-ER54109. 

\section{Declaration of Conflicts of Interest}

The authors have no conflicts of interest to disclose.

\section{Data Availability}

The data that support the findings of this study are available from the corresponding author upon reasonable request.

\bibliography{aipsamp}

\end{document}